\documentclass[aps,prl,showpacs,twocolumn]{revtex4}
\usepackage{graphics}
\usepackage[dvips]{color}
\usepackage[T1]{fontenc}
\usepackage[latin1]{inputenc}
\usepackage{graphicx}
\usepackage{amssymb}
\usepackage{rotating}
\usepackage{epsfig}

\input epsf.sty

\newcommand{\ie}{{\em i.e.}} 
\newcommand{\eg}{{\em e.g.}}
\newcommand{\etal}{{\em et al.}}

\newcommand{\dir}{{Figs}}

\begin{document}

\title{Structure of symmetric and asymmetric ``ripple'' phases in lipid
bilayers}
\author{Olaf Lenz, Friederike Schmid}
\affiliation{Fakult\"at f\"ur Physik, Universit\"at Bielefeld, D -- 33615
Bielefeld, Germany}

\begin{abstract}
We reproduce the symmetric and asymmetric ``rippled'' $P_{\beta'}$ states
of lipid membranes by Monte Carlo simulations of a coarse-grained 
molecular model for lipid-solvent mixtures. The structure and properties
compare favorably with experiments. The asymmetric ripple state 
is characterized by a periodic array of fully interdigitated ``defect'' 
lines. The symmetric ripple state maintains a bilayer structure, but is
otherwise structurally similar. The main force driving the
formation of both ripple states is
the propensity of lipid molecules with large head groups to exhibit splay.
\end{abstract}


\maketitle

Membranes are ubiquitous in all living organisms~\cite{gennis}. Their central 
structural element is a lipid bilayer, which is stabilized by the
amphiphilic character of lipid molecules -- they self-assemble such that 
their hydrophilic head groups shield the hydrophobic tails from the 
surrounding water. Pure lipid bilayers have been studied for a long time as model 
systems that can provide insight into the structural properties of biomembranes. 
Already these seemingly simple systems exhibit a rich spectrum 
of structures and phase transitions~\cite{nagle1,nagle2,koynova1,koynova2}. 
The most common state in nature is the so-called ``fluid'' state 
($L_\alpha$), which is characterized by a large number of chain defects and 
high lipid mobility. If one decreases the temperature, one encounters a phase 
transition (the ``main'' transition) to a ``gel'' state where the lipid 
molecules are more ordered and less mobile. The structure of this low 
temperature phase depends on the interactions between the lipid head groups. 
Loosely speaking, lipids with small head groups such as 
phosphatidylethanolamines~\cite{koynova1} assume a state where the long axes 
of the chains remain perpendicular to the bilayer normal ($L_\beta$ phase).
Lipids with large head groups and relatively strong head-head attraction 
such as phosphatidylcholines~\cite{koynova2} exhibit tilt ($L_{\beta'}$ phase). 
Finally, lipids with large head groups and weak head-head attraction 
such as as ether linked phospatidylcholines~\cite{koynova2,fn1} form a phase 
$L_{\beta}^{\mbox{\tiny int}}$ where both opposing lipid layers are 
fully interdigitated~\cite{koynova1,koynova2}. 

The main transition has attracted considerable interest, since it occurs at
temperatures that are typical on earth (between 
$-20^0$C and $60^0$C). The mechanism that governs the transition
$L_\alpha \leftrightarrow L_{\beta}$ to the untilted gel is comparatively 
straightforward. The transition is driven by the competition 
of the entropy of chain disorder and the free energy of chain 
alignment~\cite{schmid,whitmore} (\ie, chain packing) and is thus in some 
sense related to the isotropic-nematic transition of liquid crystals. 
At the transition $L_\alpha \leftrightarrow L_{\beta'}$ to the 
tilted gel, the situation is much more complicated. Here, the main 
transition is preempted by a ``pretransition'', and one observes an 
intermediate state with a periodic, wave-like surface structure: 
The ``ripple'' phase $P_{\beta'}$, first reported by Tardieu 
\etal~\cite{tardieu}. The microscopic structure of this mysterious 
phase has been debated for a long time. 

In fact, at least two different rippled states have been reported, which 
often coexist~\cite{tenchov}. Electron density maps (EDMs) have recently 
been derived for both of them from X-ray scattering 
data~\cite{sun,sengupta}.
One of the structures is asymmetric and has a sawtooth profile with 
alternating thin and thick arms and a periodicity of 13-15 nms, which 
corresponds to roughly 20 lipid diameters. The other one is symmetric and 
has a wavy sinusoidal structure with twice the period of the asymmetric 
structure~\cite{katsaras}. The formation of the ripples depends strongly 
on the thermal history~\cite{tenchov,katsaras,matuoka}. If the membrane
is heated up from the gel state, asymmetric ripples are obtained. If one 
cools down from the fluid state, both types of ripples are formed -- 
predominantly asymmetric ones if the cooling was fast, and predominantly 
symmetric ones if the cooling was slow and if a long time was spent at the 
transition temperature. Dynamical X-ray~\cite{katsaras} and 
AFM~\cite{kaasgard} studies suggest that the symmetric ripple state is 
metastable and very slowly transforms into the asymmetric ripple state;
however, this does not yet seem to be fully settled. The degree of ordering 
in the ripple states largely resembles that in the gel state, except for
a certain amount of disorder in the structure~\cite{nagle1} --
calorimetric studies suggest that approximately 10 \% of all chains are melted.
Most strikingly, the self-diffusion of lipids in the ripple states is a 
few orders of magnitude higher than that in the gel state, and highly 
anisotropic~\cite{schneider}. This has lead to the assumption that the 
ripple states might contain ``coexisting'' gel-state and fluid-state 
lipids. 

Numerous theoretical models for the ripple phase have been proposed,
which explain the ripple formation by different mechanisms: Chain packing
effects~\cite{larsson,pearce}, dipolar interactions~\cite{doniach}, a 
coupling of monolayer curvature with local chain melting~\cite{falkovitz,
marder,heimburg} or with tilt~\cite{larsson,carlson,lubensky} in
combination with chirality~\cite{lubensky}. This list is far from complete. 
In contrast, molecular simulations of rippled membrane states are still 
scarce. Kranenburg \etal~\cite{kranenburg} were the first to reproduce a 
periodically modulated membrane state in a dissipative-particle dynamics 
(DPD) simulation of a coarse-grained lipid model. 
They observe a periodic sequence of stripes with alternating gel 
and liquid order, similar to a structure proposed theoretically by
Falkovitz \etal~\cite{falkovitz}. Unfortunately, the distribution of head
groups in that structure is not consistent with the experimental EDMs 
-- the structure is neither asymmetric, nor does it feature the waviness 
which characterizes the symmetric ripple. Moreover, the relative fraction 
of molten molecules -- 50 \% -- seems too high, compared to experiments.
A second, very interesting simulation has recently been carried out by 
de Vries \etal~\cite{devries}. In an atomistic model of a lecithin bilayer, 
these authors found a structure containing a stretch of interdigitated 
membrane and a stretch of gel membrane. 
The interdigitated patch connects the neighboring gel membrane such that the 
upper leaflet of the bilayer on one side crosses over into the lower leaflet 
on the other side. The authors assume that this structure will repeat
itself periodically in larger systems and identify it with an asymmetric 
ripple. It is worth noting that the lipids are not arranged in a
continuous bilayer 
-- as had been assumed in all previous models for the ripple state.

In this letter, we present Monte Carlo simulations of a simplified 
coarse-grained lipid model, which reproduce asymmetric and symmetric 
ripple states with properties that compare very favorably to experiments.
The structure of the asymmetric ripple is similar to that proposed by 
de Vries \etal~\cite{devries}. Our simulations show that it is indeed
a periodic structure, and that it is generic, \ie, it does not depend
on molecular details of the lipids. Moreover, they enable us to propose 
a structural model for the symmetric ripple as well, and to identify the 
mechanisms that stabilize the rippled structures.

\begin{figure}[b]
\includegraphics[scale=0.35,angle=0]{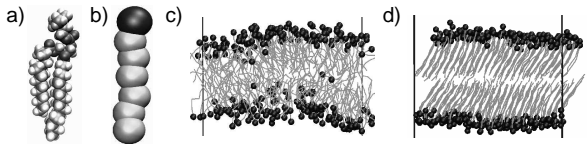}
\vspace*{-0.3cm}
\caption{\label{fig:model} 
Illustration of our lipid model and snapshots (sideview)
of two lipid states (at $\rho_s k_B T = 2 \epsilon/\sigma^3$).
(a) All-atom model of DPPC
(b) Coarse-grained bead-spring model used
in this work
(c) The fluid phase $L_\alpha$ at $k_B T=1.3 \epsilon$
(d) The tilted gel phase $L_{\beta'}$
at $k_B T=1.1 \epsilon$. 
For better visualization, only heads (reduced
size) and tail bonds are shown.
}
\end{figure}

We employ a lipid model which we have used earlier to investigate phase 
transitions in Langmuir monolayers~\cite{stadler,duechs}: 
Lipid molecules are represented by chains made of one head bead and six tail 
beads (Fig.~\ref{fig:model}b), which are connected by anharmonic springs 
and subject to an intramolecular bending potential. The tail beads attract 
one another with a truncated and shifted Lennard-Jones potential 
(diameter $\sigma$, well depth $\sim \epsilon$). The head beads are larger 
than the tail beads ($1.1 \sigma$) and purely repulsive. The other parameters 
and the exact form of the potentials can be found in Ref.~\onlinecite{duechs} 
(the model corresponding to Fig.~7). Self-assembly of the ``lipids'' is
enforced with a recently proposed ``phantom solvent'' environment~\cite{lenz}: 
We add ``solvent'' particles which interact only with lipid beads 
(repulsively), and not with one another. Two examples of self-assembled 
membranes in the $L_\alpha$ and the $L_{\beta'}$ state are shown in 
Fig.~\ref{fig:model}c,d). The phantom solvent has the simple physical 
interpretation that it probes the accessible free volume for solvent 
particles in the presence of lipids. It entropically penalizes lipid/solvent 
interfaces, and thus effectively creates an attractive ``depletion'' 
interaction between the lipid beads next to such an interface, \ie, the 
head beads~\cite{fn2}. 
The strength of this interaction is directly proportional to the phantom
solvent density, $\rho_s$.
Compared to other explicit solvent environments, the phantom solvent has
the advantage that it does not introduce an artificial solvent 
structure and artificial solvent-mediated interactions between periodic 
images of bilayers. Moreover, it is computationally cheap -- in Monte Carlo 
simulations, less than 10 \% of the total computer time is typically spent on 
the uninteresting solvent region. 

We have carried out Monte Carlo simulation at constant pressure 
with periodic boundaries in a simulation box of fluctuating size and shape.
This ensured that the membranes had vanishing surface tension.
The system sizes ranged from 288 to 1800 lipids (corresponding to 2000-12600
beads), typical run lengths were 1-10 million Monte Carlo sweeps.
\begin{figure}[t]
\centerline{
\includegraphics[scale=0.2,angle=0]{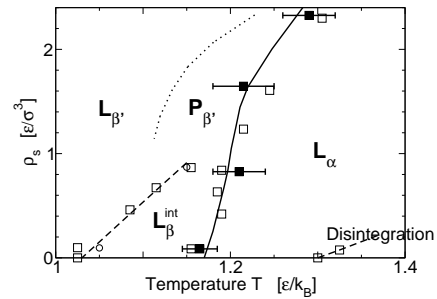}
}  
\vspace*{-0.3cm}
\caption{\label{fig:phases} 
Phase diagram of the lipid model as a function of temperature, $T$,
and phantom solvent density \ie, effective head interaction, $\rho_s$. 
The phases are: $L_\alpha$ (fluid), $L_{\beta'}$ (tilted gel), 
$L_{\beta}^{\mbox{\tiny int}}$ (interdigitated gel), $P_{\beta}'$ (ripple). 
At high temperatures, the bilayer disintegrates. Open squares indicate 
transition points from simulation runs of small systems initially set
up as fully ordered, untilted bilayers. No ripples were observed in that case.
Closed squares show transition points determined by heating up and cooling 
down the system (different system sizes), with the error bars corresponding
to the width of the hysteresis. Open circles denote points where the membrane
undergoes a transition from $L_{\beta'}$ to $L_{\beta}^{\mbox{\tiny int}}$ 
upon heating. 
}
\end{figure}
The resulting phase diagram is shown in Fig.~\ref{fig:phases}. The model 
reproduces the experimentally observed gel and fluid phases for lipids with 
large heads: The fluid phase ($L_\alpha$), the interdigitated gel 
($L_{\beta}^{\mbox{\tiny int}}$) for low $\rho_s$, \ie, weak head attraction, 
and the tilted gel ($L_{\beta'}$) for higher $\rho_s$, \ie, strong head attraction. 
The structures of these phases and the phase transitions shall be discussed 
in detail elsewhere~\cite{lenz2}. 

\begin{figure}[t]
\centerline{
\includegraphics[scale=0.25,angle=0]{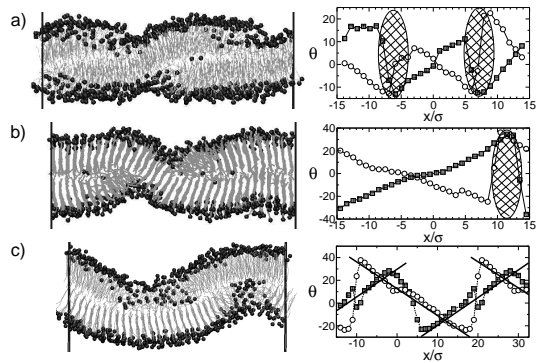}
}
\vspace*{-0.3cm}
\caption{\label{fig:ripples} 
Three examples of ripple configurations in a 
model bilayer of 720 molecules (left)
with corresponding tilt profiles $\theta(x)$ (right)
(a) Two asymmetric ripples, formed after rapidly 
cooling down to the temperature $k_B T=1.1 \epsilon$ 
from the $L_{\alpha}$ phase
(b) One asymmetric ripple, formed after heating 
up to $k_B T=1.21 \epsilon$ from the $L_{\beta'}$ phase
(c) Symmetric ripple, formed after slowly cooling down 
to the temperature $k_B T=1.18 \epsilon$ from the 
$L_{\alpha}$ phase. Note that the snapshots (left) show
sideviews of whole fluctuating configurations -- therefore, 
a large number of heads seem to be buried inside the layer, 
even though they are at the surface.
The curves in the graphs (right) are shifted in the $x$
direction and replicated periodically. Open circles correspond
to the lower monolayer, closed squares to the upper 
monolayer. The hatched ellipses indicate the regions with the 
interdigitated line defect, and the thick solid lines in c)
the slopes of $\theta$ on the ordered monolayer regions.
}
\end{figure}

Here, we fix the solvent density at $\rho_s k_B T=2 \epsilon/\sigma^3$, where 
the gel phase has the tilted $L_{\beta'}$ 
structure. In the transition region between $L_{\beta'}$ and $L_{\alpha}$, 
we observe modulated configurations which we identify with rippled states.
They develop spontaneously and reproducibly when cooling a fluid membrane
or heating a tilted gel membrane in a temperature range close to
the transition temperature. As in the experiments, their exact structure 
depends on the thermal history. Fig.~\ref{fig:ripples} (left) shows three 
examples.

The structure in Fig.~\ref{fig:ripples}a) emerged
after cooling the system rapidly from the fluid phase down to 
$k_B T = 1.1 \epsilon$. It exhibits two ripples of width $\sim 15 \sigma$
with a structure very similar to that found by de Vries \etal~\cite{devries} 
in their atomistic simulations: At each ripple, a thin interdigitated line
defect connects a lower with an upper monolayer. The second
monolayer ends at that line with an edge of disordered, melted chains. 
The period, $\sim 15 \sigma$, corresponds to the ``natural''
period of asymmetric ripples in our model. This was deduced by comparing 
simulations of systems with different sizes. The quality of the ripple formation 
depends on the initial box dimensions (before cooling). In systems of half the size, 
only one ripple formed. If one box dimension was close to a multiple of $15 \sigma$, 
the box shape readjusted to accomodate the optimal ripple width. If the initial box
dimensions were very unfavorable (20 or 24 $\sigma$), no clean ripples formed;
instead, the ripples bulged and developed interconnected structures.

The second structure, shown in Fig.~\ref{fig:ripples}b), resulted from heating
up a bilayer in the tilted gel state up to a temperature close to the 
main transition, $k_B T = 1.21 \epsilon$. During the simulation, the bilayer 
first fluctuated very strongly. After $\sim$ 6 million Monte Carlo sweeps, the 
tilt was so strong that the lipids in both monolayers slid along each other and 
connected with the other monolayer. The final structure exhibited one asymmetric 
ripple and fluctuated much less. This structure remained stable for another 
6 million sweeps. Apparently, the formation of the second ripple is prevented 
kinetically.

The third structure, Fig.~\ref{fig:ripples}c), shows a membrane 
which has been cooled down from the fluid state to a temperature close to 
the main transition $k_B T = 1.18 \epsilon$. In this case, a new type of 
structure emerged: The membrane maintains its bilayer structure, but the 
monolayers contain curved, ordered stripes with a width of roughly 25 $\sigma$. 
These ``gel'' stripes on the upper and lower monolayers are interlocked, 
such that the membrane assumes an overall sinusoidal shape. Each stripe ends 
on both sides with conical regions of disordered chains, which are very 
similar to the monolayer caps in the asymmetric ripple state. The total width 
of a ripple is $\sim 30 \sigma$, which is twice as much as the width of an 
asymmetric ripple. We identify this structure with the symmetric ripple 
state. 

\begin{figure}[b]
\includegraphics[scale=0.2,angle=0]{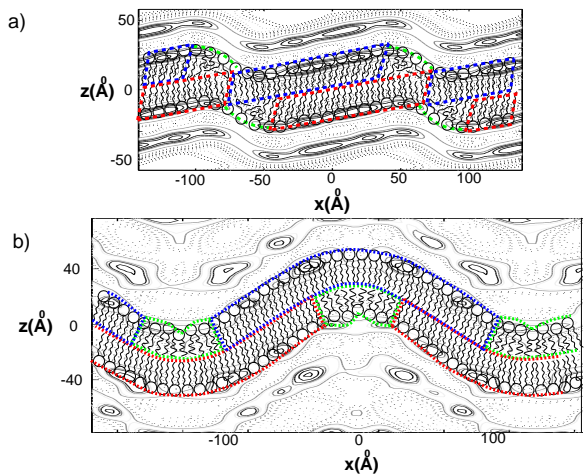}
\vspace*{-0.6cm}
\caption{\label{fig:edm} 
Sketch of the proposed microscopic structures for the
ripple states layers superimposed onto EDMs from
Ref.~\protect\onlinecite{sengupta}.
(a) asymmetric ripple (on an EDM for DMPC at 18.2 ${}^0$ C)
(b) symmetric ripple (on an EDM for DPPC at 39.2  ${}^0$ C)
}
\end{figure}

To support this hypothesis, we superimpose the proposed structures for
the asymmetric and the symmetric ripple with the EDMs of Sengupta 
\etal~\cite{sengupta} in Fig.~\ref{fig:edm}. They can be inserted
very nicely. In the asymmetric case, they explain the sawtooth shape 
with the thin and thick arm (assuming that the interdigitated region
is indeed small). In the symmetric case, they reproduce the sinusoidal 
shape. 

The simulations can be used to characterize the ripple states in more detail. 
We just summarize some of the results here, the data will be presented and 
discussed elsewhere~\cite{lenz2}. The structure of the ripple states is in many 
respect similar to that of a gel: Roughly $\sim 85\%$ of the chains have
chain lengths distributed as in the gel, only 15\% have a reduced length. 
This is in rough agreement with the experimental findings on the amount of 
chain disorder in the ripple state. The head layer in the ordered parts of the 
ripple state has the same thickness than in the gel state. The structure 
factor of the ripple state indicates a large amount of positional
order, and resembles that of an untilted gel. The most revealing
structural feature is the average tilt of the molecules. It points 
perpendicular to the ripple and is modulated. Fig.~\ref{fig:ripples} (right)
shows profiles of the average tilt angle $\theta$ for the three ripples
discussed above. The slope $\theta'(x)$ turns out to be almost constant 
throughout the whole ordered part of the monolayer. Moreover, the numerical 
values are comparable: $\theta' \sim 2.6/\sigma$ on average for the two asymmetric 
ripples, and $\theta' \sim 2.5/\sigma$ for the symmetric ripple. Even the 
single asymmetric ripple of Fig.~\ref{fig:ripples}b), which 
has an unfavorable period, still features a constant slope of
$\theta' \sim 2.3/\sigma$. This suggests strongly that the ripple
formation is primarily driven by the propensity of lipids with large head groups 
to exhibit {\em splay}. The driving force of the local curvature, \ie, head
packing, is much weaker. It does however play a role in determining
the periodicity of the ripple. Our cooling simulations clearly indicate
that the system favors a certain ripple width, for which both splay and 
curvature are optimized.

In sum, we have reproduced symmetric and asymmetric rippled states 
with a generic model for lipid membranes. The comparison with experiments is
favorable: The structure is consistent with the available EDMs, the
period length is of the same order as the experimental period 
length ($\sim 15$ lipid diameters), the amount of chain disorder is comparable, 
and we observe the same dependence on thermal history. Therefore, we believe 
to have strong evidence that our structures correspond to the real ripple 
states observed in experiments. Factors that are important for the formation 
of these states are: (i) The vicinity to the $L_{\alpha}$ phase, such 
that a small number of chains can melt, (ii) a strong tendency of monolayers 
to splay inwards -- caused by a mismatch between head group and tail size, 
and (iii) the possibility to interdigitate. 
Chirality is not necessary, in agreement with experiments~\cite{katsaras}; 
the ``lipids'' do not even have to be asymmetric. 

In fact, the factor (iii) is only needed to stabilize the asymmetric ripple 
state. If it is absent, the system can still form a symmetric ripple state. 
We note that the symmetric and the asymmetric ripples are structurally quite 
similar. Both contain about the same amount of molten chains, both have 
large ordered monolayer regions with comparable splay. This explains why 
the two types of ripples coexist, and why it seems so hard to determine 
which one is stable. Our results suggest that the answer to that question 
may depend on the type of lipid, \eg, on the head interactions and other 
factors that promote or prevent interdigitation. Coarse-grained lipid models 
may help to study this systematically. Unfortunately, we have not yet been 
able to develop an efficient strategy to determine the free energy difference 
between the two states. This will be subject of future work.

We thank V.A. Raghunathan for providing us with the EDMs,
and the NIC computer center J\"ulich for computer time. 
This work has been funded by the Deutsche Forschungsgemeinschaft
within the SFB 613.

\clearpage


\begin{thebibliography}{66}

\bibitem{gennis} R. B. Gennis, {\em Biomembranes}, Springer Verlag (1989).

\bibitem{nagle1} J. F. Nagle, Ann. Rev. Phys. Chem. {\bf 31}, 157 (1980).

\bibitem{nagle2} J. F. Nagle, S. Tristram-Nagle, 
  Biochim. Biophys. Acta {\bf 1469}, 159 (2000).

\bibitem{koynova1} R. Koynova, M. Caffrey,
  Chem. Phys. Lipids {\bf 69}, 1 (1994).

\bibitem{koynova2} R. Koynova, M. Caffrey, 
  Biophys. Biochim. Acta {\bf 1376}, 91 (1998).

\bibitem{fn1} In general, lipids are ester linked. By changing the chain
linkage type from ester to ether, one removes a strong hydrogen bond
acceptor, which reduces the tendency of head groups to form hydrogen
bonds with one another.

\bibitem{schmid} F. Schmid, M. Schick, J. Chem. Phys. {\bf 102}, 2080 (1995).

\bibitem{whitmore} M. D. Whitmore, J. P. Whitehead, A. Roberge,
 Can. J. Phys. {\bf 76}, 831 (1998).

\bibitem{tardieu} A. Tardieu, V. Luzzati, F. C. Reman, 
 J. Mol. Biol. {\bf 75}, 711 (1973).

\bibitem{tenchov} B. G. Tenchov, H. Yao, I. Hatta, Biophys. J. {\bf 56},
757 (1989).

\bibitem{sun} W. J. Sun \etal, PNAS {\bf 93}, 7008 (1996).

\bibitem{sengupta} K. Sengupta, V. A. Raghunathan, J. Katsaras,
  Phys. Rev. E {\bf 68}, 031710 (2003).

\bibitem{katsaras} J. Katsaras \etal, Phys. Rev. E {\bf 61}, 5668 (2000).

\bibitem{matuoka} S. Matuoka \etal, Biophys. J. {\bf 64}, 1456 (1993).

\bibitem{kaasgard} T. Kaasgaard \etal, Biophys. J. {\bf 85}, 350 (2003).

\bibitem{schneider} M. B. Schneider, W. K. Chan, W. W. Webb,
  Biophys. J. {\bf 43}, 157 (1983).

\bibitem{larsson} K. Larsson, Chem. Phys. Lipids {\bf 20}, 225 (1977).

\bibitem{pearce} P. A. Pearce, H. L. Scott, J. Chem. Phys. {\bf 77}, 951
(1982).

\bibitem{doniach} S. Doniach, J. Chem. Phys. {\bf 70}, 4587 (1979).

\bibitem{falkovitz} M. S. Falkovitz \etal, PNAS {\bf 79}, 3918 (1982).

\bibitem{marder} M .Marder \etal, PNAS {\bf 81}, 6559 (1984).

\bibitem{heimburg} T. Heimburg, Biophys. J. {\bf 78}, 1154 (2000).

\bibitem{carlson} J. M. Carlson, J. P. Sethna, Phys. Rev. A {\bf 36},
3359 (1987).

\bibitem{lubensky} T. C. Lubensky, F. C. MacKintosh,
 Phys. Rev. Lett. {\bf 71}, 1565 (1993).

\bibitem{kranenburg} M. Kranenburg, C. Laforge, B. Smit, 
  Phys. Chem. Chem. Phys. {\bf 6}, 4531 (2004).

\bibitem{devries} A. H. de Vries \etal, PNAS {\bf 102}, 5392 (2005).

\bibitem{stadler} C. Stadler, H. Lange, F. Schmid,
  Phys. Rev. E {\bf 59}, 4248 (1999).

\bibitem{duechs} D. D\"uchs, F. Schmid,   
  J. Phys.: Cond. Matt. {\bf 13}, 4853 (2001).

\bibitem{lenz} O. Lenz, F. Schmid,
  J. Mol. Liquids {\bf 117}, 147 (2005).

\bibitem{fn2} Free head beads and solvent demix at $\rho_s \sim 2.6/\sigma^3$.

\bibitem{lenz2} O. Lenz, F. Schmid, in preparation.

\end{thebibliography}
\end{document}